# The Planets in Indigenous Australian Traditions

Duane W. Hamacher[1,2] and Kirsten Banks[3]

[1]Monash Indigenous Studies Centre, Monash University, Clayton, VIC, Australia
[2]Centre for Astrophysics, University of Southern Queensland, Toowoomba, QLD, Australia
[3]School of Physics, University of New South Wales, Sydney, NSW, Australia

**Emails**: duane.hamacher@gmail.com | kirstenabanks.kb@gmail.com

## Abstract

Studies in Australian Indigenous astronomical knowledge reveal few accounts of the visible planets in the sky. However, what information we do have tells us that Aboriginal people were close observers of planets and their motions, noting the relative brightness of the planets, their motions along the ecliptic, retrograde motion, the relationship between Venus and its proximity to the Sun, Venus' connection to the Sun through zodiacal light, and the synodic cycle of Venus, particularly as it transitions from the Evening Star to the Morning Star. The dearth of descriptions of planets in Aboriginal traditions may be due to the gross incompleteness of recorded astronomical traditions, and of ethnographic bias and misidentification in the anthropological record. Ethnographic fieldwork with Aboriginal and Torres Strait Islander communities is revealing new, previously unrecorded knowledge about the planets and their related phenomena.

## 1      Introduction

The study of Indigenous astronomy has revealed a wealth of information about the role of the Sun, Moon, stars, and planets in the traditional Knowledge Systems of Aboriginal Australians (Norris 2016). Indigenous cultures around the world, particularly the many hundreds that exist in Australia, maintain complex astronomical knowledge systems that link the positions and motions of celestial objects and to navigation, calendars, subsistence, and social applications (Hamacher 2012; Johnson 1998; Cairns and Harney 2003).

Indigenous people are careful observers of subtle changes in the positions and properties of celestial objects and noted both obvious and rare celestial phenomena, such as meteors (Hamacher and Norris 2010), comets (Hamacher and Norris 2011a), eclipses (Hamacher and Norris 2011b), fireballs (Hamacher *et al.* 2017), impact events (Hamacher and Norris 2009; Hamacher 2013a), meteorite cratering (Hamacher and Goldsmith 2013), pulsating variable stars (Hamacher 2017), aurorae (Hamacher 2013b), stellar scintillation (Guedes *et al.* 2017), the analemma (Norris et al. 2013), lunar phases (Hamacher and Norris 2011c), novae (Hamacher and Frew, 2010), and supernovae (Hamacher 2014).

The role of the planets in Indigenous Australian cultures (both Aboriginal and Torres Strait) is a topic that has not been researched in detail (Fredrick 2008). Aboriginal traditions provide insight to ancient knowledge about planets and planetary motion and could help modern researchers in both the sciences and social sciences in their work, focusing on mutual benefits.

The most detailed literature-based study of the planets in Aboriginal traditions was completed by Fredrick (2008). Although it missed several key references, her thesis provides an informative basis on which we build a further study. This paper presents a preliminary study showing how Indigenous Australians perceived, conceptualised, and applied knowledge of planets and their motions, including the ecliptic, retrograde motion, synodic cycles, the relationship of planets to other celestial objects, zodiacal light, and solar and lunar dynamics.

## 2    The Planets in Indigenous Australian Traditions

The Indigenous people of Australia include the Aboriginal people of mainland Australia and Tasmania, and the Torres Strait Islanders, who inhabit the archipelago between Cape York and Papua New Guinea. Archaeological evidence places Aboriginal people in Australia (or the larger landmass of Sahul) around 65,000 years ago (Clarkson *et al.* 2017). There are over 500 recorded languages in across Australia, each with a unique culture and traditions (McConnell and Thieberger 2001). Thus, Indigenous views of planets and planetary phenomena are diverse and widespread.

The Sun, Moon, and visible planets (Mercury, Venus, Mars, Jupiter, and Saturn) were known to Aboriginal and Torres Strait Islander people. These cultures paid careful attention to the motions of solar system bodies through careful observation, which was recorded and passed to successive generations through oral tradition and material culture. Aboriginal and Islander people distinguished planets from the background stars, noted their changing positions in the sky, their changing positions relative to each other, their proximity to each other along the zodiac of the ecliptic, and their dynamic relationship to the Sun and Moon.

Descriptions of planets in Aboriginal and Torres Strait Islander cultures are scant in the literature compared to other astronomical objects and phenomena. Venus features prominently, but planets such as Mercury, Mars, Jupiter, and Saturn are less common. Fredrick (2008: 34) collated 526 traditional stories about astronomy, of which 4% included Venus and (3%) included the remaining planets. The most prominent objects discussed are the Moon (16%), Orion and the Pleiades (11%), and the Milky Way (8%).

Given the plethora of misidentifications, misinterpretations and conflated identifications in ethnographic research related to Australian Indigenous astronomical knowledge, these omissions are likely the result of the ethnographers than a lack of Indigenous traditions about these celestial bodies. Research on the subject is ongoing, but this article provides a current overview on the role of planets and related phenomena in Aboriginal and Torres Strait Islander traditions.

In most Aboriginal cultures (but not all), the Moon is a man and the Sun is a woman - often spouses (Clarke, 2009; Johnson, 1998). The planets are often seen as prominent sky ancestors and direct relations to the Sun and Moon. These relationships vary from romantic to familial relationships – and even rival relationships. To the Tiwi people, the Moon man follows the path of the Sun woman and has four wives who also follow this path. They are represented by Mars, Mercury, Venus and Jupiter (Mountford 1958). To the Kamilaroi people, Jupiter is a young boy who is disliked by his mother, the Sun. She sends men to spear the boy when he is low in the western sky (Tindale, 1983).

Some planets are viewed as hierarchical figures. Venus – typically represented as both the Morning and Evening Star – for example, is seen as a senior spirit and was the boss of all of the other stars (Mathews and Dixon 1975). In Boorong traditions of western Victoria, Venus is *Chargee Gnowee*, sister of the Sun (*Gnowee*) and wife of Jupiter (*Ginabongbearp*). Ginabongbearp is chief of the the old spirits (*Nurrumbunguttias*) (Stanbridge 1857).

Planets can also be associated with other entities, such as totems or even emotions. Venus, the Evening Star, is a woman associated with a big lizard totem (Spencer and Gillen 1927). In Kamilaroi traditions, Venus is an old man who once said something rude and has been laughing at his own joke ever since (Parker 1905).

## 3    The Path of the Sun, Moon, and Planets

The *ecliptic* traces the Sun's apparent path around the Earth and is the basis of the ecliptic coordinate system. The planets orbit the Sun in roughly the same plane and are inclined by approximately 9° from the ecliptic. The region within 9° either side of the ecliptic is the *zodiac*. The path of the Sun Moon, and planets is widely known across many Aboriginal regions. It is generally seen as a road or pathway for the primary ancestor spirits, sometimes as a road to the land of the dead on which spirits travel to the afterworld (Cahir *et al.*, 2018). Wardaman people see the zodiac as a road that ancestor spirits use to travel across the sky and is utilised for navigation (Cairns and Harney 2003). In Tiwi traditions of Melville and Bathurst Islands, north of Darwin, the Sun woman carries a torch across the sky each day, from East to West (describing diurnal motion). The Moon man follows the path of the Sun woman but carries a smaller torch. Four planets - Mars, Mercury, Jupiter, and Venus - represent the Moon man's four wives, who also travel across this pathway (Mountford 1958: 174). In the Western Desert, amateur anthropologist Daisy Bates (1936) noted that some Aboriginal communities perceived Jupiter and Venus "always following one another along the 'dream road' which they themselves had made." Jupiter and Venus are visible during the day, which may contribute to their close relationship.

Some Aboriginal traditions describe the motions of planetary ancestor spirits as they walk along the zodiacal road in the sky. *Retrograde motion* is a visual effect describing the motion of a planet as it travels through the night sky each night while taking into account the relative position of the planet with respect to the background stars. Derived from the Latin word *retrogrades* (meaning backwards-step), it describes the apparent backward motion of a planet. Generally, planets appear to move from the East to the West. When a planet is in retrograde motion, it appears to slow to a stop (relative to the background stars each night) before moving backwards from West to East (Fig 1). All of the planets in the Solar System have periods of retrograde motion. The Wardaman people articulate this phenomenon clearly by describing the planets as the old spirits who would walk the path, both forwards and backwards (Cairns and Harney 2003: 64-65).

> "The Dreaming Track in the sky! Planets making the pathway! Travelling routes, a pathway you could call it, like a highway! Travelling pathway joins to all different areas, to base place, to camping place, to ceremony place, where the trade routes come in; all this sort of things. The Dreaming Track in the sky, the planets come straight across ... walking trail becomes a pad, then becomes a wagon road, two wheel tracks, then become a highway. That's how they started off."

The roles of the planets themselves are varied and diverse. They are related to totems and marriage classes, as well as ancestral figures in the sky who inform various social practices and traditional laws. For example, the Moon and Venus are subordinate totems of *Nyaui*, the Sun clan, in western Victoria (Mathews, 1904). Planets are distinguished from other stars by their movement with respect to background stars, their tendency not to twinkle (e.g. Haddon 1912: 219), as well as the pathway they travel across the sky. Traditions of individual planets, bar Venus, are scant in the ethnographic literature.

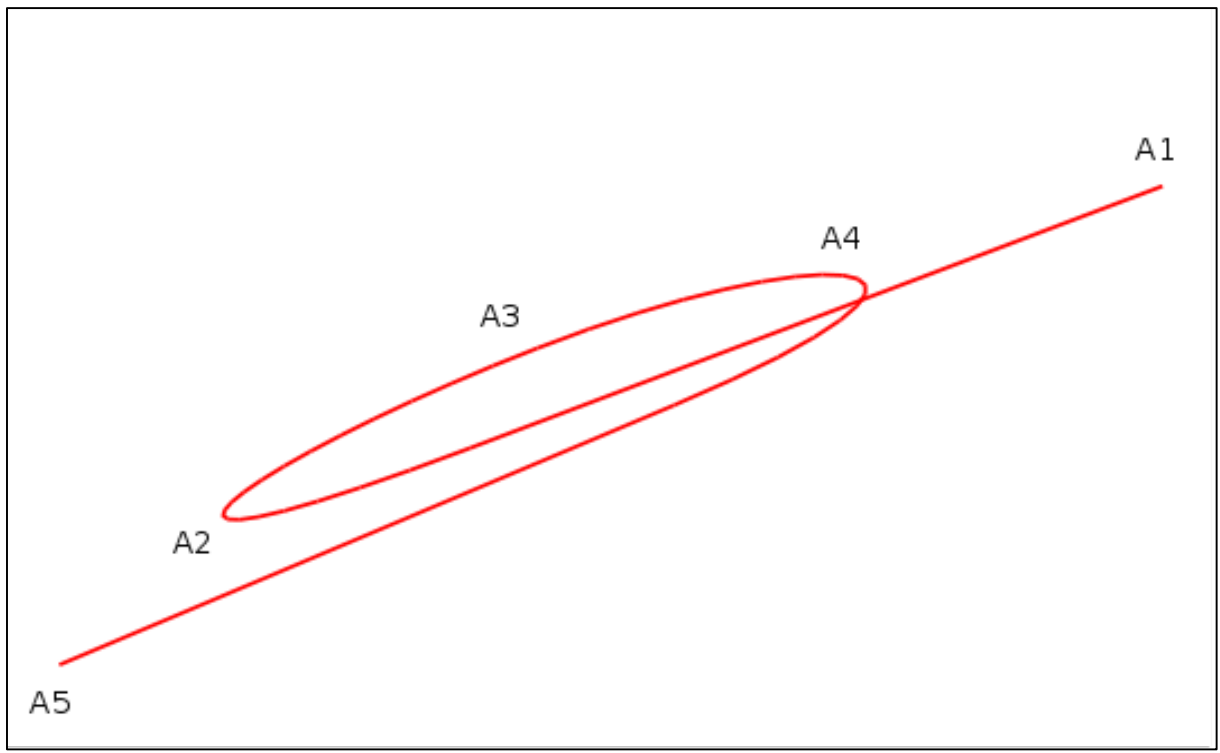

**Fig. 1**. *Image displaying retrograde motion of Mars over the course of several months. Image: Wikimedia Commons.*

## 4 The Planets

### 4.1 Mercury

Very few recorded Aboriginal traditions speak directly about the role of Mercury. The Wardaman people describe the planet as a little girl named *Gowaman* that relates to the damaging actions of the Moon-man (Cairns & Harney 2003: 62-65). In Kamilaroi traditions, Mercury probably represents a red kangaroo that relates to ceremony (Fuller, 2015: 123). Elders working with Fuller (2015: 108) said that "the red kangaroo was a star low in the western sky after sunset, and it was very important in ceremony." The elder said the 'star' was red, difficult to see, and only there "some of the time" – suggesting Mercury as its likely identity. The red colour of the planet is probably due to reddening caused from its low altitude near the horizon, but the identity of Mercury as the object in question is inferred – not explicitly identified.

### 4.2    Venus

As the third brightest object in the sky, Venus is the most commonly described planet in Aboriginal traditions. It is often described as the Morning and Evening star. Traditions of Venus are too numerous to include in this article, but they often possess social meaning, rather than subsistence or calendric application, due to its wandering nature in the sky. Given its position as an inferior planet (closer to the Sun than the Earth), it is commonly linked directly to the Sun and Moon.

In Kamilaroi and Euahlayi traditions, Venus (particularly when it is low on the horizon) is seen as an old man who is laughing animatedly after telling a rude joke (Parker 1905: 71). These same communities also speak of the Eaglehawk, *Mullyan*, who lived in a large yarran tree. He hunted people for food near the Barwon River. In response to his killing of the people, a group of young men set his home alight, killing him. The Eaglehawk then ascended into the sky as *Mullyangah*, the Morning Star (Parker 1898: 31-32; Reed 1965: 79-82). The story is reflected in the landscape at waterholes bearing the name and story at Morgan's Wells near the Glengarry opal fields of northern New South Wales (Fuller 2015). Other places feature Venus as the Morning Star, such as *Kabminye* - a place near Tanunda (north-east of Adelaide) (Cockburn, 1908 [1984: 111]).

In Murrawarri traditions of western New South Wales, Venus (Mirnkabuli) is a man who lives in a hut (gurli), living on mussels and crayfish (Mathews 1904: 283). The Kamilaroi and Euahlayi people also describe Venus and Mars as the eyes of the celestial being Baayami (Fuller 2015). During special ceremonies held near Quilpie in western Queensland, the Euahlayi met with the Arrernte of the MacDonnell Ranges in Central Australia. The Arrernte bring a red stone, signifying Mars, and the Euahlayi people bring a green and blue stone, representing Venus. With some initial confusion about the nature and role of Venus/Morning Star in relation to the Eaglehawk and the eyes of Baayami, one of Fuller's (2015: 67) consultants clarified that:

> *"during the day, Maliyan's eyes (the eaglehawk) are the eyes of Baayami. During the night, Maliyan's eyes are Venus and Mars, which become the eyes of Baayami at night. Because one is red, and one is blue and green, two rocks are brought together for ceremonies: one is red (opal) from Quilpie, QLD, and the Euahlayi have the green and blue one (opal). When the stones are put together, they are Venus and Mars on Earth. Another location for Venus and Mars on Earth is a place called Mordale near Narran Lakes. These are waterholes side by side, called Milmaliyan and Maliyan-ga."*

Venus is also connected to sacred ceremony in these communities. According to Fuller's Aboriginal consultants, the Evening Star serves as a sign to light the sacred fire. The fire is relit every evening until the Morning Star is seen, at which time the ceremony takes place (and the sacred fire is doused). This caused confusion, as Venus cannot be seen together as a Morning and Evening Star. An elder explained to Fuller (2015) that the Morning Star in this case was actually Mars, again tying the two planets together as the eyes of Baayami.

In Yolngu traditions of Arnhem Land in the Northern Territory, traditions of Venus as the Morning Star are commonplace and serve a major role in ceremony (Clarke 2014). In some Yolngu communities, Venus is perceived as a spirit taking the form of a lotus flower atop a stem representing its path (Berndt, 1948). A related song (ibid: 35) explains that Venus is:

> *"Shining on to the fore-heads of all those head-men. On to the heads of all those Sandfly [clan] people. It sinks into the place of the white gum trees, at Milingimbi."*

The Aboriginal people of the Hamilton and Georgina Rivers in Queensland referred to Venus as *Mumungooma*, or "big-eye" (White 1904; Clarke 2014). They perceive Venus as fertile grasslands with an abundance of seeds used to make flour. This land, they believed had no water, but if the Aboriginal people who inhabited this land grew thirsty, they could travel to Earth via ropes hanging from it. These ropes are reflected in the Banimbirr ceremony, which links white strings connecting Venus to the Sun (Fig. 2).

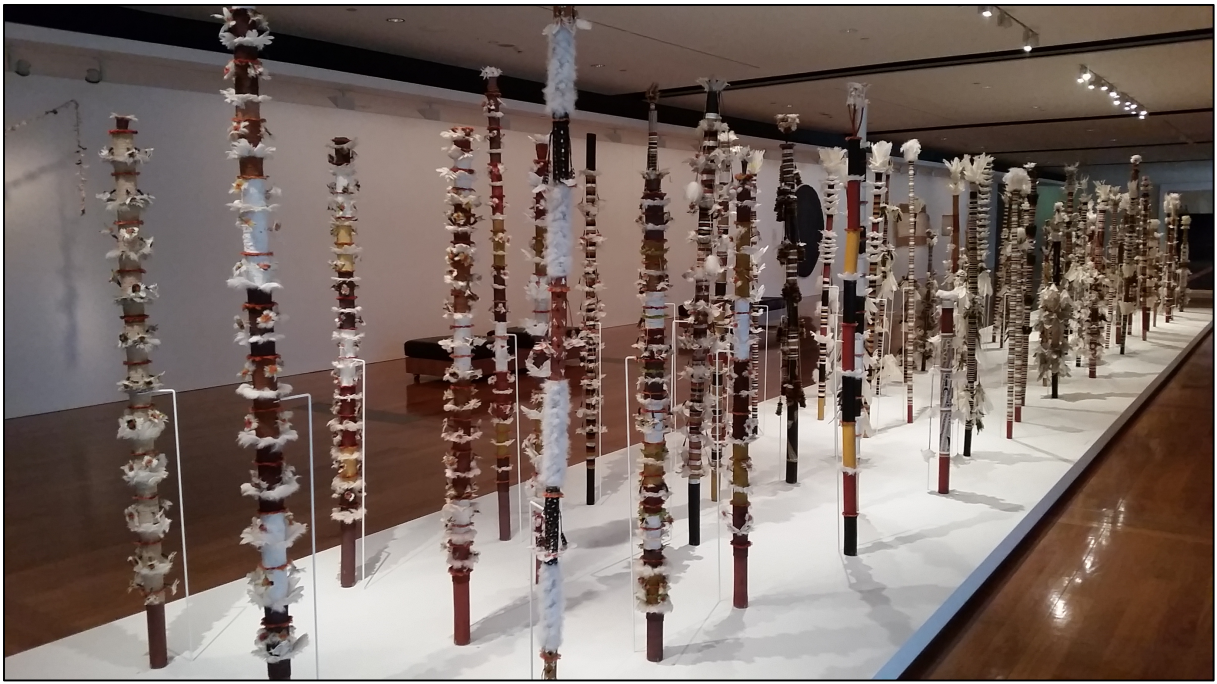

***Fig. 2****. Banumbirr Morning Star poles from Aboriginal artists on Galiwin'ku (Elcho Island), Northern Territory displayed at the Gallery of Modern Art in Brisbane. The white feathers symbolise starlight and the symbolic power of shining. Photo: Author Supplied.*

Aboriginal and Torres Strait Islander people note the distinction between Venus as the Morning Star and the Evening Star. They are seen as the same object, but visible at different times (never the same time). The Yolngu, as well as other communities, hold the Banumbirr ceremony when Venus first rises (at a particular azimuth on the horizon) after transitioning from the Evening Star to the Morning Star (e.g. Norris and Norris 2009: 18-22). Yolngu people hold the Banumbirr ceremony to observe Venus rising between the mainland and Burralku - the sacred island of the dead – to the east as it makes this transition. The ceremony starts at dusk and continues through the night, reaching a climax when Banumbirr rises a few hours before dawn (Norris and Norris 2009). Banumbirr communicates with the people through a faint rope that also keeps her close to the Sun. When Norris and Norris (2009) asked a Yolngu elder how they knew when to hold the Morning Star ceremony, the elder said they simply count the days.

This indicates the Yolngu people were aware – at least to some degree - of the synodic period of Venus. Because Venus and the Earth are in 5:8 resonance, if one were to record the position of Venus each night at the same time they would discover that the planet produces five unique 'patterns' that repeat every eight years. The first rising of the Morning Star as it transitions from an Evening Star occurs approximately every 584 days – a phenomenon closely observed by the Maya of Central America (Aveni 1997: 45). Aboriginal people in other parts of Australia were also careful observers of this phenomenon and kept close track of when the ceremony was to be held (Fuller 2015).

The Morning Star Pole contains symbolism of a light rope that connects Venus with the Sun, representing a pathway by which spirits can travel. This is probably a reference to zodiacal light, which is easily seen from the dark skies of the region. As an inferior planet, Venus does not stray far from the Sun and is often visible in the bright band of zodiacal light, which is a reflection of sunlight off of interplanetary dust in the Solar System.

Venus has a tighter orbit around the Sun compared to the Earth's orbit. Due to this, the planet is often seen close to the Sun in the sky, either trailing it or leading it. Venus is often referred to in modern astronomy as the Morning Star or the Evening Star, depending on what time of day it is visible. This description was also adopted by Indigenous Australians. There are many accounts of Venus being described as the Morning Star and the Evening Star. Both terms were often used to refer to the same object, meaning Indigenous Australians knew that both the Morning Star and the Evening Star are the same object, namely, Venus.

At Tanunda, north-east of Adelaide, the Aboriginal people describe Venus as Kabminye, meaning Morning Star (Cockburn 1908). In the Gulf of Carpentaria, Venus is recorded to be both the Morning and Evening Stars (record 297). It is explained as,

> *"the main female character (as Venus) was both the Morning and Evening Star. She was promised to the sharkman, but did not want to marry him so escaped into the sea, only to be followed by the sharkman. Desperate not to be caught by the man she flew up into the sky and became the Evening Star, the sharkman can still be seen sometimes in the waters around Morington Island."*

In the Torres Strait, Venus is given two titles. In the Kala Lagow Ya language of the western islands, the Morning Star is *Goeyga Thithuuyi* and the Evening Star is *Woey* (Eseli 1998). In the Merima Mir language of the eastern Torres Strait, the Evening Star is Ilwel and the Morning Star is Gerger Nasau (Hamacher 2015).

The Morning and Evening Star are sometimes assigned a social function. For example, it is included in the Arrernte tradition about Tnorala - a ring-shaped mountain range 5 km across and 150 m high in the Central Desert 160 km west of Alice Springs. The Western Arrernte tradition is that in the creation period, a group of women took the form of stars and danced a corroboree (ceremony) in the Milky Way. One of the women was carrying her baby, so she put him in a wooden basket (turna) and set him down near the edge of the Milky Way. As the women danced the corroboree, they shook the Milky Way and the baby fell off. He came tumbling down as a star and struck the ground with the turna covering him, driving the rocks around him upwards to create the ring-shaped mountain range.

Tnorala, called Gosses Bluff by Western scientists, is the remnant of a highly eroded complex impact crater, formed when a low-density projectile (probably a comet) struck the Earth around 142 million years ago. The original crater is roughly 22 km wide, with a 5km wide central uplift. At the time of impact, the land was nearly 2 km higher. Tnorala is the highly eroded central uplift of the original crater. In Arrernte traditions, the turna can still be seen in the sky – tumbling from the Milky Way – as the Western constellation Corona Australis (the Southern Crown). The baby's parents – the Morning and Evening Star – take turns searching for their lost child to this day. Parents warn their children not to stare at the Morning or Evening Star, as the baby's celestial parents might mistake the gazing child as their lost offspring and whisk them away to the sky (Thornton 2007).

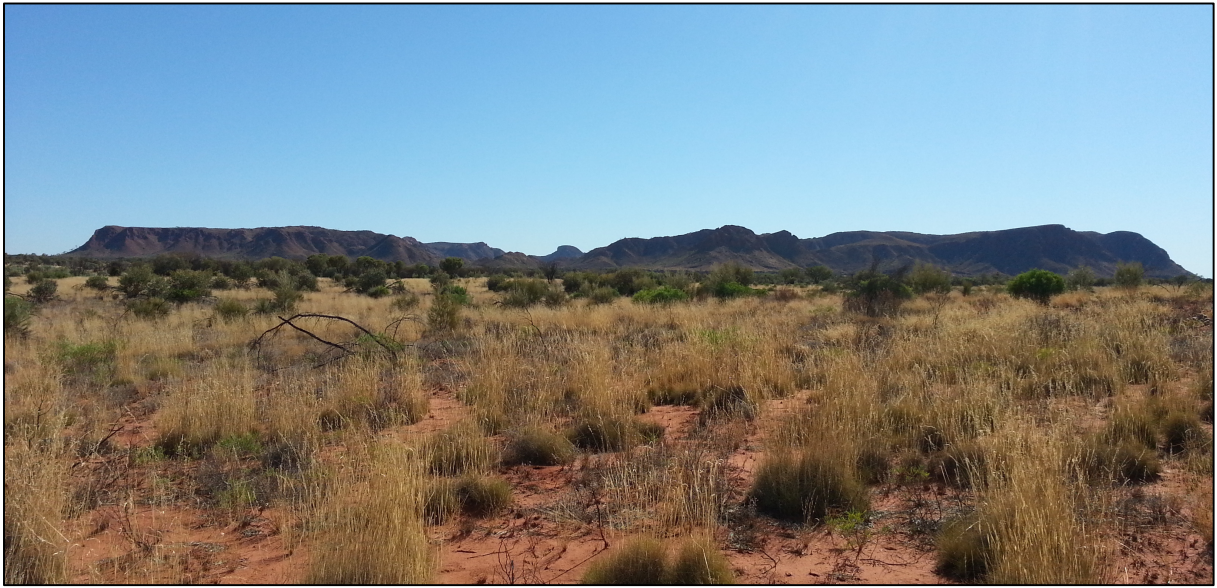

**Fig. 3.** *Tnorala (Gosses Bluff crater) in the Central Desert. Photo: Author supplied.*

An additional link between Venus and its proximity to other celestial bodies is found in the waning crescent Moon's link to Venus as the Morning Star. In the Gulf country of North Queensland (Hulley 1996; Isaacs 1980; Reed 1993), two brothers travel across the land in the Creation period. The elder brother is the Moon and the younger brother is the Morning Star. They use a boomerang to form landscape features, such as rivers, valleys, hills, and seas. As they move inland, the bush is increasingly thick. They throw their boomerang to the West, clearing a path in front of them from which the ocean tide follows. As the brothers rest that night, the elder Moon brother pretends to sleep, then awakens when his younger brother falls asleep. The Moon brother uses his boomerang to cut off the genitals of his brother and moulds him a pair of breasts, turning him into a woman. When the younger brother awakens, he finds he is now the wife of the Moon brother. The sighting of the crescent Moon is now linked closely with the Morning or Evening Star (whether the Moon is a waxing or waning crescent) and serves as a reminder to the people. Given Venus' inferior orbit to Earth, the Moon will always be in a crescent phase when seen near Venus, as either a Morning or Evening Star. The story also informs the people about the diurnal motion of the celestial objects each night and the link between Moon phases and ocean tides (Leaman 2012).

### 4.3    Mars

Mars appears in the traditions of a number of Aboriginal cultures across Australia, but is overall poorly represented in the ethnographic literature. The Kamilaroi people refer to it as "gumba", meaning "fat" (Ridley 1888). The ruddy colour of Mars held special significance. The Anmatyerre-speaking people of the Central Desert described it as *lherrm-penh*, meaning "something that has been burnt in flames" (Green, 2010: 395). The Kokatha people of the Western Desert associate Mars and Antares with *Kogolongo*, the red tailed black cockatoo (Leaman and Hamacher 2014; Fig 4). This is reminiscent of the Classical traditions of these objects, as Mars is the Roman god of war and Ares is the Greek god of war. Antares means "equal to Ares." Since both stars are of similar brightness and colour, and since they occasionally come close to each other in the sky, they are seen as rivals.

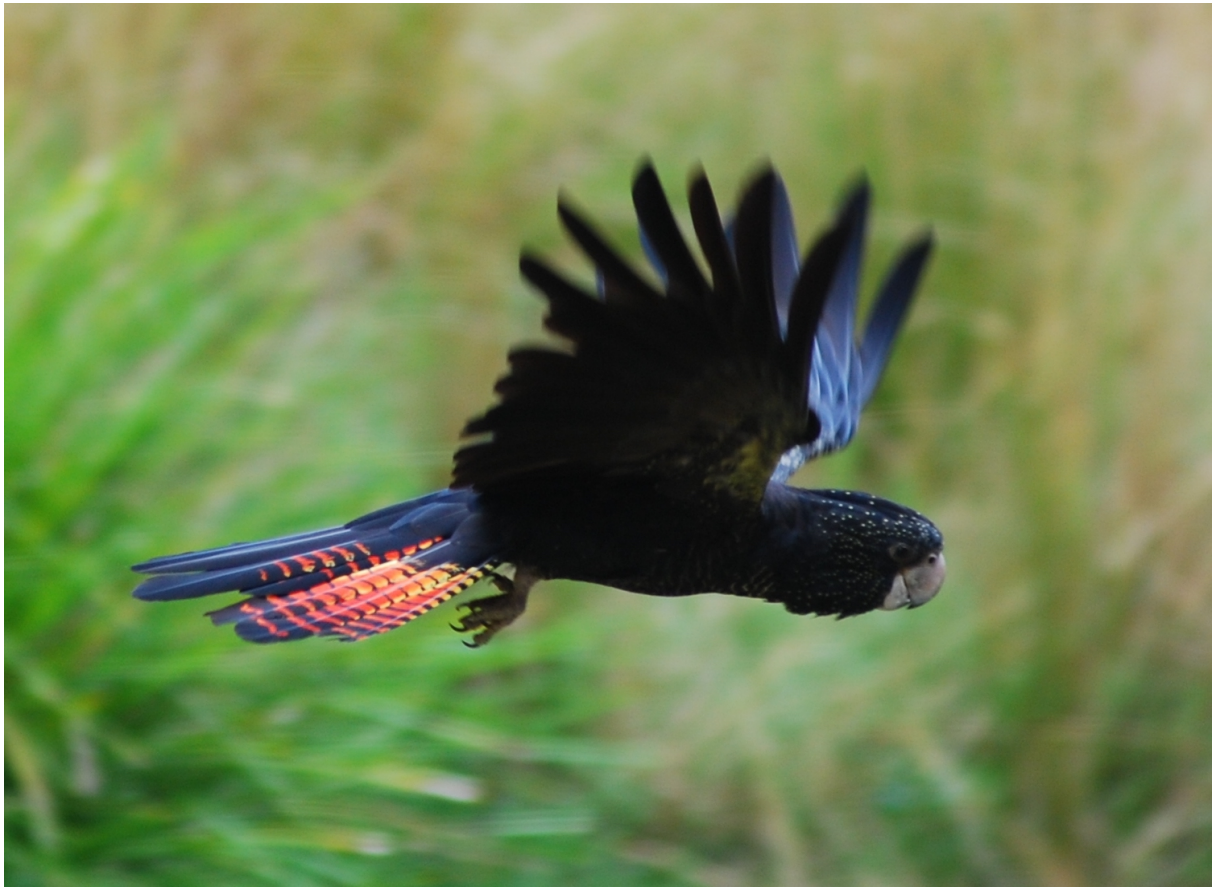

**Fig 4.** *A red-tailed black cockatoo, related to both Mars and Antares in Kokatha traditions. Image: Peter Campbell via Wikimedia license.*

The Aboriginal people of Oyster Bay, Tasmania hold a tradition that two ancestral men stood on a mountaintop and "threw fire, like a star … [that] fell among the black men" (Milligan 1859: 274). The pair of men live in the clouds and can be seen as the Greek Gemini twins, Castor and Pollux. In August 1831, Tasmanian man Mannalargenna told Englishman George Augustus Robinson that the men (Pumpermehowlle and Pineterrinner) created fire and now live in the skyworld. "*Mars was his [sic] foot and the Milky Way his [sic] road"* (Robinson and Plomley, 2008: 872). As seen from Australia, the orientation of the Greek Gemini twins is upside down, yet this is the description given by the Tasmanian men: Pumpermehowlle and Pineterrinner appear to walk along the pathway of the Milky Way. As their foot, Mars would appear in the region between the two stars and the plane of the Milky Way. Earlier, in May 1831, Mars was visible between Castor and Pollux and the Milky Way, which might explain the association. It is peculiar that Mars would be seen as the foot, since Mars wanders the heavens and will only occasionally be in this orientation (Gantevoort *et al*. 2016).

Anthropologists recorded a tradition from the Coorong south of Adelaide about a young man going through an initiation ritual named Waiyungari, meaning "red man", since he is covered in red ochre during the ritual period of time (Clarke 1999). He engages in illicit behaviour with two women, breaking a traditional law. To escape punishment, his casts a spear into the Milky Way and pulls himself and the women into the sky. Waiyungari is a red star with the two women flanking him on either side in the celestial canoe (the stars of Scorpius). For 150 years, anthropologists misidentified the red Waiyungari as Mars, and suggested the two women were other planets. Hamacher (2017) re-analysed the traditions and realised that Waiyungari is actually the star Antares (Alpha Scorpii) and the the women are Tau and Sigma Scorpii – not Mars, or any other planet.

### 4.4    Jupiter

Jupiter is prominent in the sky but has few identified traditions associated with it in the ethnographic record. Like Mars, it is commonly associated with ruddy colours. Aboriginal people of the Darling River, NSW (probably the Murrawarri, see Fuller 2015) call Jupiter *Wurnda wurnda yarroa*, an ancestral man who subsided on roasted yams. The ruddy colour of the planet is a reflection of the fire used to roast the yams (Mathews 1904: 283).

The wandering motion of Jupiter, like other planets, was noted by many Aboriginal groups. In northern NSW, the Euahlayi and Kamilaroi share a story about the planet (Fraser 1888: 8-9):

> *"In the grasslands of the eastern riverine corridor west of the Great Dividing Range, peoples of several tribes have stories based on the idea that Jupiter is a young boy wandering about the heavens. He is much disliked by his mother, the Sun, so much so that she sends men to spear him at a time when he is moving low down in the western sky. In general the fear of people is that in dry years the grasses may not set seed, and if the Sun woman succeeds in injuring her son this will be sure to happen. An even greater fear is that if the boy were "killed" all people would become ill, would develop blindness, and many would perish. Even Kukura, their Moon man, could go blind. Such ideas appear to reflect their own experiences with drought and with the effects of severe malnutrition caused thereby."*

A Kamilaroi man from the Lake Coocoran area, NSW told Fuller *et al.* (2014: 21) that Jupiter was a

> *"red-eye fella. Kids don't play with fire; red-eye fella will follow you and stay all winter."*

He described this as an admonition to kids not to play with the campfire. Another Kamilaroi/Euahlayi person told Fuller et al. that their grandmother taught them Jupiter was a wandering spirit and a bad person:

> *"that big star is watching you. Bad spirit. Do you if you're alone."*

In the Ooldea region of South Australia's Great Victoria Desert, amateur anthropologist Daisy Bates (1904-1912) recorded that Jupiter and Venus were two men travelling along their Dreaming tracks with only heads and no body. She also recorded that Jupiter was a man who assisted two young men who fed him after they were attacked by a mob. Bates states that Jupiter went up into the sky, with the two boys close by. She does not reveal the celestial identity of the young boys in the sky.

In Wangkamana traditions of central-west Queensland, a son and his mother were travelling across country and became separated, eventually ascending to the sky as Jupiter (the son) and an unnamed star (the mother) (Fraser 1901). A neighbouring community wanted to kill the son, but were unable to. The mother and son's community are greatly afraid the son will be killed. This will result in them being unable to collect food and they would starve. The recorded story is incomplete and fragmented, leaving us with this unclear and somewhat confusing account.

In these traditions, Jupiter is associated with other stars in the sky, but they are not identified by their Western counterpart. It is unclear if these "stars" are indeed stars or other planets.

### 4.5    Saturn

Like Mercury, Jupiter, and Mars, stories about Saturn are minimal in the recorded literature. The planet is associated with a small bird (*wunygal*) by the Kamilaroi and Wailwan people (Ridley 1875: 141). In the Western Desert, the local people viewed Venus (Iruwanja) and Saturn (Irukulpinja) as brothers, with Jupiter as their dog. Irukulpinja and the dog spend most of their time catching food for Iruwanja (Mountford 1976).

**Acknowledgements**

We acknowledge the traditional knowledge of the Aboriginal and Torres Strait Islander communities featured in this paper, and pay our respects to elders and custodians of the past, present, and future. We thank Dr Edwin Krupp for nominating us to write this overview and the editor and referees who provided critical and useful feedback. Hamacher is funded by Australian Research Council project DE140101600. Ethnographic fieldwork in the Torres Strait was conducted under Monash Human Research Ethics project HC15035.

**ABOUT THE AUTHORS**

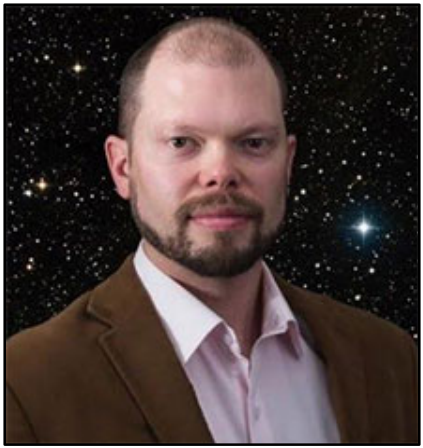 **Dr Duane W. Hamacher** is an astronomer and Senior ARC Discovery Early Career Research Fellow at the Monash Indigenous Studies Centre and an Adjunct Fellow in the Centre for Astrophysics at the University of Southern Queensland. His work focuses on astronomical heritage and Indigenous astronomical knowledge and traditions in Australia and the Pacific. He is Secretary of the *International Society for Archaeoastronomy and Astronomy in Culture*, Chairs the International Astronomical Union's Working Group on Intangible Heritage, serves on the IAU Working Group for Star Names, and is an Associate Editor of the *Journal of Astronomical History and Heritage*.

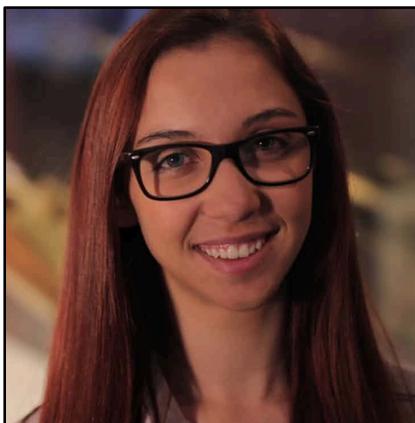 **Kirsten Banks** is a Wiradjuri woman, astronomy educator, and science communicator in Sydney, Australia. She earned a degree in astrophysics from the University of New South Wales and works as an astronomy educator at Sydney Observatory. She completed an astrophysics student internship at the Australian Astronomical Observatory and also conducts research on Australian Aboriginal astronomy. She has appeared in numerous media, including COSMOS Magazine and a major Japanese documentary that was viewed by over 10 million people. Kirsten has written for *The Guardian* and the *Australian Night Sky Guide*, and has presented at numerous astronomy events and conferences across Australia. She publishes *The Skyentists* podcast with Dr Angel Lopez-Sanchez.